\documentclass{aa}
\usepackage{graphicx}
\usepackage{natbib}
\usepackage{color}
\newcommand{\tw}{\color{black}}

\begin{document}
\headnote{}
\title{Links between magnetic fields and plasma flows in a coronal hole}


\author{T. Wiegelmann, L.D. Xia$^1$, and E. Marsch}

\institute{Max-Planck-Institut f\"ur Sonnensystemforschung,
Max-Planck-Strasse 2, 37191 Katlenburg-Lindau, Germany\\
$^1$School of Earth and Space Sciences, Univ. of Science and Technology of China, Hefei 23006, China}

\offprints{T. Wiegelmann \\ \email{wiegelmann@linmpi.mpg.de}}


\date{DOI: 10.1051/0004-6361:200500029 \\
Astronomy and Astrophysics, Volume 432, Issue 1, March II 2005, pp.L1-L4 }

\abstract{We compare the small-scale features visible in the
Ne~{\sc{viii}} Doppler-shift map of an equatorial coronal hole (CH)  as observed by
SUMER with the small-scale structures of the magnetic field as constructed
from a simultaneous photospheric magnetogram by a potential magnetic-field
extrapolation. The combined data set is analysed with respect to the
small-scale flows of coronal matter, which means that the Ne~{\sc{viii}} Doppler-shift
used as tracer of the plasma flow is investigated in close connection with
the ambient magnetic field. Some small closed-field regions in this largely open CH
are also found in the coronal volume considered. The Doppler-shift
patterns are found to be clearly linked with the field topology.
}

\titlerunning{Fields and flows in coronal hole}
\authorrunning{Wiegelmann et al.}

\maketitle

\section{Introduction}
In this study we present observations of an equatorial coronal hole (CH) and
investigate the magnetic field structures in the source regions of the fast
solar wind. We directly compare the Dopplergrams obtained by the SUMER (Solar
Ultraviolet Measurements of Emitted Radiation) instrument, yielding the
plasma flow velocity, with the photospheric magnetograms at the bottom of the
CH, which via magnetic field extrapolation gives us the
coronal magnetic field. In the case studied here we thus combine
spectroscopic data with the three dimensional (3-D) magnetic field, an approach
providing a clearer physical picture of the plasma conditions and flow
pattern prevailing in the corona. This CH was studied before by
\cite{xia03} and \cite{xia2003}. Therefore we refer the reader to their papers
for further details.

The new aspect here is the 3-D magnetic field, which is
constructed from magnetograms and extrapolated to all heights above the
photosphere. We use the Greens function method to compute the current-free
magnetic field from the scalar potential,
\begin{eqnarray}
\Phi({\bf r}) = -\frac{1}{2 \pi} \int_{\partial \Omega} B_z({\bf r^{'}})
\, \frac{d \sigma^{'}}{|{\bf r}-{\bf r^{'}}|}, \;\; {\bf B} =  \nabla \Phi,
\end{eqnarray}
where $B_z$ is the line-of-sight (LOS) photospheric magnetic field, $\partial
\Omega$ denotes the bottom boundary surface (photosphere) of the computational box,
$d \sigma = dx \, dy$, ${\bf r}=\sqrt{x^2+y^2+z^2}$, and ${\bf B}({\bf r})$ is the
3-D magnetic field.
{\tw Details of the extrapolation
technique for CHs and the quiet Sun have been described in earlier work by
\cite{solanki2004}.} Potential fields can be reconstructed from the LOS component of
the photospheric magnetic field alone. The corresponding measurements are available,
e.g., from the MDI (Michelson Doppler Imager) LOS magnetograph data on SOHO.

For the reconstruction of the dynamic coronal magnetic field, such as in twisted loops,
it is usually necessary to include the currents, which often are assumed to be
parallel to the magnetic field. These force-free magnetic fields are mathematically
more difficult to model, and one needs for their reconstruction additional
observational input data, such as the photospheric magnetic field vector
\citep{wiegelmann2004},
or optical images of the coronal plasma
structures. This method has been used by \cite{marsch2004}, who carried out a
comprehensive study of fields and flows in active regions (ARs)
associated with sunspots.
\begin{figure*}
\setlength{\unitlength}{1.0cm}
\begin{picture}(12,6)
\put(-0.7,0){\includegraphics[width=12cm]{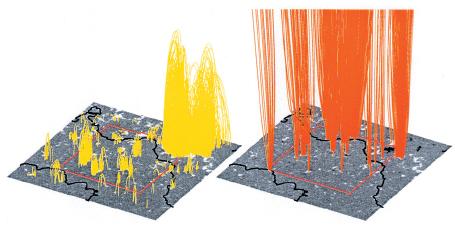}}
\end{picture}
\parbox{5.5cm}{\vspace{-4.5cm}\caption{Magnetic structures in a coronal hole.
The gray-coding shows the
field strength in the photosphere. The black line gives roughly the
boundary of the coronal hole. {\tw The field of view for SUMER is marked as a red
rectangle in both panels.} The magnetic field was constructed from a MDI magnetogram. Left
figure: mostly closed loops at various scales.
Only closed magnetic field lines with $B \ge 30$~G are shown. Right figure: Only open
fields with large photospheric values, $B \ge 100$~G. The open flux is bundled in
narrow uniform filaments and originates in stronger fields concentrated at small-scale
footpoints.
The flux tubes expand as they extend into the corona. The average expansion
factor (ratio of the maxium to the minimum of the flux tube area) is
$28 \pm 11$.}
\label{fig1}}
\end{figure*}
\section{Data analysis and field extrapolation}
During the first week of November in 1999, the SUMER telescope was pointed
to a large equatorial coronal hole and traced its darkest parts with raster
scans. The selected spectral window was centered around 154~nm, a range which
includes the lines of Si~{\sc{ii}} (153.3~nm), C~{\sc{iv}} (154.8~nm) and
(155~nm), and Ne~{\sc{viii}} (77~nm) in 2nd order. Their formation temperatures
span from about $1.8 \times 10^4$~K to $6.3 \times 10^5$~K.
The lines are emitted in the upper chromosphere, transition region and lower
corona, respectively.

The SUMER data selected for this study were taken on 5 November 1999 and analysed by
\cite{xia2003}.
The CH discussed here is shown in their Fig.1.
The instrument and data analysis techniques are described in detail
by \cite{wilhelm1997}. The detector A and slit 2 with a size of 1\arcsec $\times$
300\arcsec~ were used. SUMER began to raster in the East-West direction at 470\arcsec E
and ended at 185\arcsec E (solar disk coordinates). The corresponding times are
17:07~UT and 21:07~UT, respectively. A 150~s exposure time and 3\arcsec~ raster
step size were selected. The slit's center was pointed at 100\arcsec N during the
whole scan. The total data set consists of 96 exposures.
Here we use a magnetogram from MDI after \cite{scherrer1995}.
The MDI magnetogram, taken at 19:11 UT, has a spatial resolution of
1.98\arcsec $\times$1.98\arcsec~
and was coaligned with the SUMER images.
{\tw We used the magnetogram with $230 \times 254$ pixels
and extrapolated the magnetic field into the corona up to a
vertical height of $98$\arcsec~ with the help of the Greens-function method. The magnetic
field lines shown in Fig. \ref{fig1} correspond to a box of $454$\arcsec~ in $x$,
$502$\arcsec~ in $y$ and $50$\arcsec~ in $z$, where $z$ is the height above the
photosphere. The field lines are stretched in $z$-direction, and we cut off the picture
at a height of $50$\arcsec~ to make the small closed loops visible. The aspect ratio
between the vertical and horizontal dimension is 9 to 1.}

The results of the extrapolation are illustrated in Fig. \ref{fig1}. Various
magnetic field structures in the corona are obvious. The gray-coded bottom plot
shows the weak unipolar magnetic field strength in the photosphere, as derived from
data of the MDI instrument. The black line delineates the boundary of the coronal
hole, as identified in the EUV images obtained by EIT (Extreme Ultraviolet Imaging
Telescope) and SUMER on SOHO. {\tw  Magnetic field lines are treated as
open when they reach the upper boundary of our computational box ($98$\arcsec).
From these extrapolations alone we cannot tell definitely whether the identified open
field lines are globally open or close somewhere else on the Sun. Global potential
field source surface models have been used to reproduce the location of
CHs, but these models do not provide their fine structures and
have a source surface (around $2.5 R_{\rm S}$)
at which the field lines are forced to be open. A comparison of full-Sun EIT-images with
global coronal magnetic field in particular for CHs
was made by \cite{solanki2004b}. This work shows in its Fig. 1 that field lines
in spectroscopically identified coronal holes that are open at $100$\arcsec~
do remain open beyond 2 solar radii to the source surface. }

The closed magnetic field lines (yellow) in the left frame
of Fig. \ref{fig1} pertain to
photospheric fields with strength $B \ge 30$~G, the open ones (red-brown)
in the right frame to stronger fields, with $B \ge 100$~G. Very few field
lines are separately shown, to not confuse the picture by intermingling of
magnetic field lines, which would produce a diffuse unresolved pattern.

Note that the open magnetic flux (right frame) is concentrated, with
linear bundles reaching the top of the simulation box. These bundles are
anchored in small (a few seconds of arc in diameter) tubes linked to fine
structures with strong unipolar photospheric magnetic flux. Despite their
filamentary nature, the flux tubes spread with height, and as coronal funnels
fill increasingly larger fractions of the CH at greater altitude.

Not unexpectedly, in the CH there are only low-lying loops. They sparsely
populate the hole area, consistent with the statistical results of
\cite{solanki2004}.
Note, in contrast, that the strong
loops just outside of the hole reach greater heights. This region also
has much brighter ultraviolet emission due to a better plasma
confinement and higher plasma density.

The corresponding radiance maps, Ne~{\sc{viii}} Dopplershifts
and magnetograms are shown in Fig.~1 and Fig.~2
of \cite{xia2003}.
Here a complementary magnetogram obtained from NSO/Kitt Peak (\cite{jones1992}),
is overlaid with contours of the Doppler shifts in km~s$^{-1}$ of the
Ne~{\sc{viii}} (77~nm) line.

\begin{figure*}
\includegraphics[clip, width=12cm]{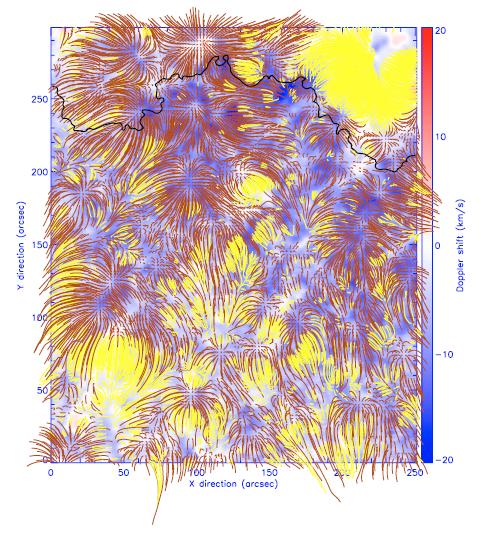}   
\parbox{5.5cm}{\vspace*{-15cm}\caption{Map of the Doppler shift of
the Ne~{\sc{viii}} (77~nm) emission line, together with the projections of the
extrapolated coronal magnetic field lines. Here yellow means closed and red-brown
open field lines. An area of 300\arcsec $\times$ 250\arcsec~ on the solar disk is
shown, covering a large fraction of the equatorial CH and its boundary regions. Note
the uniformly yellow domain in the top right corner, which lies outside the hole and
corresponds to closed loops. The cross-shaped spines of open flux in the top left
corner coincide with bluish patches and thus indicate sizable  coronal
plasma outflow in this open field domain, in particular at location
(100\arcsec,~ 250\arcsec) {\tw inside the CH. Outside the CH, e.g. at location
(110\arcsec,~ 290\arcsec) we find an open field region
 with hardly any blueshift.}
Note further that there are many open field lines overarching the yellow patches,
corresponding to the closed magnetic carpet with flux confined to near the
solar surface.} \label{fig2}}
\end{figure*}

In Fig. \ref{fig2} we present the same Dopplermap,
with the Doppler shift  over the range of $\pm 20$~km~s$^{-1}$, and
the extrapolated magnetic field lines shown in projection for
comparison with the Dopplershift pattern. Again the closed field lines are
yellow and the open ones red-brown.
An area of $250\arcsec \times
300\arcsec$ on the solar disk is shown, covering a large fraction of the equatorial
CH and its boundary regions.
Note the uniformly yellow domain in the top right
corner, which lies outside the hole and corresponds to closed loops shown before in
Fig. \ref{fig1}. The cross-shaped spines of open flux, in the top left corner for
instance, mostly coincide with bluish patches and thus indicate sizable
plasma outflow in this open field domains, in particular at location
(100\arcsec,~ 250\arcsec). Furthermore, note that there are many open field lines
overarching the smaller yellow patches, which correspond to the
closed magnetic carpet.

Detailed inspection of Fig. \ref{fig1}, and a comparison with the
previously published figures in \cite{xia2003}, indicates that almost
everywhere in the present figure do we find Doppler blue shifts, i.e.
plasma outflow along the line of sight at speeds of typically up to
10~km~s$^{-1}$. There are hardly any significant redshifts. In the yellow
domain, with closed magnetic flux, the plasma is nearly at rest.

{\tw The reader should be
aware of alternative explanations of Doppler shifts in terms of wave motion instead
of mass flow. For example redshifts in transition region lines were explained as
indications of wave motion by \cite{1994AdSpR..14...57H},
 and blueshifts cannot unequivocally be interpreted
as signs of outflow, but may also be related to upward propagating waves. }

\begin{figure}
\begin{center}
\setlength{\unitlength}{1.0cm}
\begin{picture}(8,6)

\put(-0.7,-0.7){\includegraphics[clip,
width=9.0cm,angle=0]{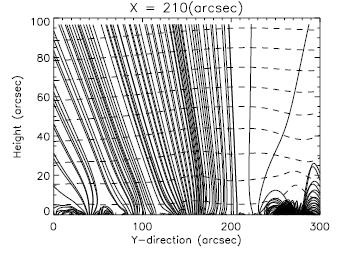}}
\end{picture}
\end{center}
\caption{Magnetic field lines projected on a cut through the extrapolation box
in the $y$-$z$-plane for the horizontal coordinate
$x=210$\arcsec. The small-scale loops at the bottom of the domain tend to push
overlaying weaker fields upwards. Note some strongly expanding funnels
at $y=40$\arcsec~ at low heights (below about $10$\arcsec). We see that the height
of closed loops in the hole (for $x<200$\arcsec)  is below about $5$\arcsec.
Outside the hole ($x>200$\arcsec) we find closed loops up to a height of about
$30$\arcsec.} \label{fig3}
\end{figure}

In order to visualize better the fine structures of the magnetic field as a function
of height (coordinate $z$) in the corona, we present in Fig. \ref{fig3} a cut
through the box in the $y$-$z$-plane at $x=210$\arcsec . This presentation
facilitates the clear identification of open and closed (in the simulation box)
field lines.
The small-scale loops of the magnetic carpet at the bottom of the domain tend to
push overarching weaker fields upwards. The overlaid horizontal lines give the
level of constant magnetic field pressure (magnetic magnitude squared) to indicate
whether the field is uniformly stratified, or if pressure imbalances do exist.
However, these spatial variations
rapidly decline with height, and above about $20$\arcsec~ the overall coronal field
is more uniform, and horizontally varies merely on larger scales of the size of
$100$\arcsec.

Close inspection of the previous Fig. \ref{fig2} shows that the strongest
blue shifts are found in the rosette-type rapidly expanding magnetic funnels. These
are the sources of coronal plasma outflow and supply of plasma in the CH to the
outer corona and solar wind.
 In regions with
closed magnetic loops (both inside and outside the CH) we do not generally
observe blueshifts.
Several papers (see, e.g., \cite{hassler1999},
\cite{wilhelm2000}, \cite{xiama03}) have addressed the issue of where the
solar wind originates, however all previous analyses relied solely on Doppler maps
that were grossly correlated with photospheric magnetograms and/or network images
obtained in cool chromospheric ultraviolet emission lines. Following the AR study of
\cite{marsch2004}, this is a first attempt to correlate in a CH the plasma flow
pattern with the 3-D magnetic field structure.

\section{Conclusions}
The present study has emphasized the need to analyze solar ultraviolet images and
spectrogramms or Doppler-shift maps in close connection with the coronal magnetic
field, which can routinely be constructed and extrapolated to the outer corona from
photospheric magnetograms. The field is the key player in the low-beta corona in
constraining plasma flow and guiding the nascent solar wind outflow through open
coronal funnels, which according to Fig. \ref{fig3}  have a more complicated
geometry at low heights (below about $10$\arcsec) than assumed in the models
considered by \cite{marsch1997} or \cite{hackenberg2000}.
The flux tube geometry in the transition region may be of importance to the initial
acceleration of the solar wind (see, e.g., \citeauthor{2002ApJ...566..562L}
\citeyear{2002ApJ...566..562L}).

The results in Fig. \ref{fig2} are of central importance to appreciate the
key role played by the coronal magnetic field. They corroborate previous
findings of SUMER, and complement the results published
by \cite{xia2003} and in the thesis of \cite{xia03}.
The Ne~{\sc{viii}} Doppler-shift maps show a close relationship with
the magnetic carpet structure and funnels inside the CH. Largest blue shifts
with speeds up to 20~km~s$^{-1}$ (darkest blue patches) are associated with
those regions where intense open magnetic fields of a uniform polarity are
concentrated. In contrast, the plasma confined in small loops in the CH does
not reveal any significant flow.

\begin{acknowledgements}
The work of Wiegelmann was supported by  DLR-grant 50 OC 0007. SUMER and MDI are
part of SOHO of ESA and NASA.  We thank the referee V.~H. Hansteen for useful
remarks. We thank D. Markiewicz-Innes for usefull comments.
\end{acknowledgements}

\end{document}